\documentclass[fleqn,twoside]{article}
\usepackage{espcrc2}
\usepackage{amssymb}

\usepackage{graphicx}
\usepackage[figuresright]{rotating}

\newcommand{\be}{\begin{equation}}
\newcommand{\ee}{\end{equation}}
\newcommand{\bea}{\begin{eqnarray}}
\newcommand{\eea}{\end{eqnarray}}
\newcommand{\ep}{\epsilon}

\newcommand{\mpi}{M_{\pi}}
\newcommand{\fpi}{F_{\pi}}

\newcommand{\mr}{\mathrm}

\newcommand{\MeV}{\,\mr{MeV}}

\newcommand{\fm}{\,\mr{fm}}

\def\fs{\; \; .}
\def\co{\; \; ,}

\newcommand{\AmS}{{\protect\the\textfont2
  A\kern-.1667em\lower.5ex\hbox{M}\kern-.125emS}}

\title{Finite volume effects in chiral perturbation theory}

\author{Gilberto Colangelo\address{Institut f\"ur Theoretische Physik der
Universit\"at Bern \\
Sidlerstrasse 5, 3012 Bern, Switzerland}}

\begin{document}

\begin{abstract}
There has recently been an intense activity in the study of finite volume
effects by means of chiral perturbation theory. In this contribution I
review recent work in this field both for the $\epsilon$-- ($M_\pi L
\lesssim 1$) and the $p$--regime ($M_\pi L \gg 1$). For the latter I
emphasize the importance of going beyond leading order calculations in
chiral perturbation theory and the usefulness of asymptotic formulae {\em
  \`a la} L\"uscher used in combination with CHPT.  \vspace{1pc}
\end{abstract}

\maketitle

\section{Introduction}
Every lattice calculation is done in finite volume and at finite lattice
spacing, as well as at finite (and at present typically substantial) quark
masses. Before being able to make a meaningful comparison to experimentally
measured quantities, one has to make three extrapolations -- and since all
of them are numerically quite expensive, any analytical method that may
help in this respect is more than welcome. Chiral perturbation theory
(CHPT) provides the proper framework for calculating analytically the
dependence on all three extrapolation parameters. An overview of the
present status of the analytical calculations related to the extrapolation
to the continuum and to the chiral limit, and in particular of the
interplay between the two limits, has been given by Oliver B\"ar
\cite{Baer}. Here I will discuss the extrapolation to the infinite volume
limit, and give an overview of the present status of analytical
calculations in the framework of CHPT.

If a physical system is enclosed in a finite box, and periodic boundary
conditions are imposed, the space of the possible three-momenta becomes
discrete:
\be
\vec{p} = \frac{2 \pi}{L} \vec{n}, \qquad \vec{n}=(n_1,n_2,n_3)
\ee
with $n_{1,2,3} \in \mathbf{Z}$. If the box is large enough, this discrete
space will almost look like a continuum one: and since one is interested in
simulating the soft, nonperturbative dynamics of QCD, one would like to
have the region of soft momenta to look like a continuum one. If we take 
$4 \pi F_\pi$ as the chiral symmetry breaking scale, which separates soft
from hard momenta, we obtain the following quantitative condition on the
volume
\be
\frac{2 \pi}{L} \ll 4 \pi F_\pi \qquad \Rightarrow \qquad L \gg \frac{1}{2
  F_\pi} \sim 1 \fm 
\label{eq:LFpi}
\ee
in order to have a large number and not only a few discrete momenta in the
soft region.

If the condition (\ref{eq:LFpi}) is satisfied one can then use CHPT to
study analytically the behaviour of the system at low momenta, and in
particular the explicit dependence of physical observables on the
volume. How to extend the CHPT framework to the finite volume case has been
discussed by Gasser and Leutwyler \cite{GLFV}: in finite volume CHPT
becomes a systematic expansion in both the quark masses and the inverse box
size. In infinite volume there are relations among the coefficients of this
expansion for different quantities which follow from the chiral symmetry of
QCD: these relations go traditionally under the name of ``low energy
theorems'' and CHPT is a convenient tool to derive them
systematically. In finite volume chiral symmetry implies relations among
the coefficients of the expansion in $1/L$ of different observables as well
as relations among these coefficients and those of the expansion in quark
masses. CHPT allows one to obtain these ``large volume theorems'' in a
systematic way. 

The product $F_\pi L$ is not the only relevant parameter. An important role
is also played by the relative size of $M_\pi$ and $1/L$.  If $M_\pi L
\lesssim 1$ finite volume effects will be similar to those in the chiral
limit -- as is well known, no spontaneous symmetry breaking can take place
in a finite volume, and one will see a deformation of the vacuum state due
to finite volume effects. If $M_\pi L \gg 1$, on the other hand, the effect
of the explicit symmetry breaking on the low energy behaviour of the system
will be more important than that due to the finite volume. In the
framework of CHPT, the difference between the two situations is given by
the importance that the Goldstone-boson zero modes have in the evaluation
of the path integral. In the latter case, like in infinite volume, the
contribution of the zero modes can be neglected, whereas in the first case
is important and must be explicitly evaluated \cite{GLFV,HL}. The first
situation is denoted as ``$\ep$--regime'' and the second one
``$p$--regime'', and the respective counting schemes are as follows: 
\bea 
\epsilon\mbox{--regime:} && M_\pi
\sim \frac{1}{L^2} \sim O(\epsilon^2) \nonumber \\ p\mbox{--regime:} &&
M_\pi \sim \frac{1}{L} \sim O(p) \fs 
\eea

In the first case the Compton wavelength of the pion is larger than the box
size, and the pion does not have enough space to propagate (all other
non-Goldstone-like particles however do, because their mass is of the order
of $\Lambda \sim 4 \pi F_\pi$, and we assume that the condition
(\ref{eq:LFpi}) is satisfied). In the latter case, a pion fits comfortably
well inside a box and, being the lightest particle, is the only one that
has a nonnegligible probability to propagate until the box boundaries, and
therefore feel their presence. 
\begin{figure}[t]
\includegraphics[width=78mm]{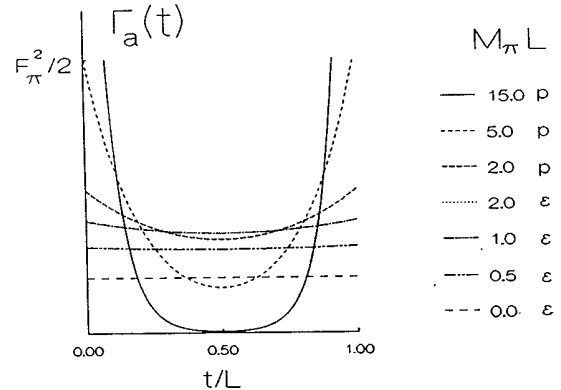}
\caption{Time-dependence of the axial-charge correlator on for different
  values of $M_\pi L$. Note that for $M_\pi L=2$ the calculation in the
  $\ep$-- and $p$--regime give perfectly overlapping results. Figure taken
  from \protect\cite{HL} 
\label{fig:AAHL}}
\end{figure}
The different behaviour of the physical
system in the two cases is well illustrated in Fig.~\ref{fig:AAHL} where
the time dependence of the correlator of two axial charges
\be
\langle Q_i^A(t) Q_k^A(0) \rangle \equiv \delta_{ik} L^3 \Gamma_A(t)
\label{eq:AA}
\ee
is plotted for different values of $M_\pi L$. In infinite volume the
correlator drops exponentially with a rate which is proportional to
$M_\pi$:
\be
\Gamma_A^{L=\infty}(t)= M_\pi F_\pi^2/2 \; e^{-M_\pi t}
\left[1+O(p^4) \right] \fs
\ee
In the $p$--regime one expects a behaviour of the correlator which is
qualitatively similar to the one in infinite volume -- as
Fig.~\ref{fig:AAHL} indeed shows. In the $\ep$--regime, on the other 
hand, the qualitative behaviour is completely different from the
infinite-volume case -- the curves in Fig.~\ref{fig:AAHL} corresponding
to small values of $M_\pi L$ can be well described by a polynomial of low
degree in $t$ \cite{HL}. 

\section{Finite volume effects in the $\ep$--regime}
One of the goals of lattice calculations is to determine the low-energy
constants of the CHPT Lagrangian. These constants are by definition
independent of the light quark masses and a reliable determination should
therefore be made with quark masses as small as possible, which is
notoriously expensive. Even when very small masses will become
accessible\footnote{In the quenched approximation very small quark masses
  have already been simulated, see e.g. \cite{Lchiral}.} working close to
the chiral limit {\em and} in the $p$--regime will require enormous volumes
and will likely remain for a long time prohibitively expensive. Approaching
the chiral limit on the lattice {\em implies}  working in the $\ep$--regime.
As discussed above, observable quantities may look very different in the
$\ep$--regime from the infinite-volume case. The strategy of the
calculation in this case must start from an analytical calculation in CHPT
to identify the proper observable that will allow an extraction of the
low-energy constant of interest. The first analytical calculations in this
direction have been made in \cite{HL,LS}. The first numerical calculations
which exploited this strategy were aimed at the extraction of the quark
condensate in the chiral limit \cite{Lqq,Lchiral}.

More recently, two different groups engaged in the calculation of
charge-charge correlators of the type defined in Eq.~(\ref{eq:AA}) in order
to extract the pion decay constant in the chiral limit. Results have been
published during last year \cite{QQ1,QQ2} and show a very clean
determination of this chiral-limit observable. In fact one of 
the two groups has performed the calculation of the decay constant also in
the $p$--regime and has shown that after the
extrapolation to the chiral limit the result is in agreement with the
direct determination in the $\ep$--regime:
\bea
F &\simeq& 130 \; \MeV \hskip 4.47 cm \mbox{\cite{QQ1}}  \nonumber \\
F &=& (102 \pm 4 \; \MeV )_\epsilon \co \; \; \; (104 \pm 2 \; \MeV)_p 
\quad \mbox{\cite{QQ2}} \nonumber
\eea
Although the calculations have been made in the quenched approximation, the
positive results prove the feasibility of the method and have opened the way
for a more ambitious aim of the program: the calculation of the low-energy
constant of the weak chiral Lagrangian for nonleptonic $K$-decays. The
first results in this direction have been presented at this conference 
\cite{Wittig} and have also recently appeared \cite{WLEC}.

Unfortunately, these groups have also found out that working in the
$\ep$--regime is technically challenging, and to complete their program had
to overcome unexpected, highly nontrivial difficulties of numerical
nature \cite{QQ1,QQ2,Wittig}. It will be interesting to see the application
of this strategy with dynamical fermions -- which is certainly a highly
nontrivial task. 

Two papers have recently investigated this regime for the one-nucleon
sector and promise interesting developments \cite{epsnuc}.

\section{Finite volume effects in the $p$--regime}
Gasser and Leutwyler \cite{GLFV} have shown that, if one works in the
$p$--regime, in a large isotropic box with periodic boundary conditions,
the only change in the calculation is that the propagator of the pseudo
Goldstone bosons has to be made periodic:
\be
G_L(\vec{x},t)= \sum_{\vec{n}} G_\infty(\vec{x}+\vec{n}L,t) \fs
\label{eq:GL}
\ee
The local vertices, as given by the CHPT Lagrangian, remain unchanged and
the power counting for loop diagrams works like in infinite volume: every
loop generates an additional $O(p^2)$ correction. The only difference 
is that in finite volume $p \sim M \sim 1/L$, and a one-loop calculation
gives both the leading quark-mass and finite volume corrections.
For example, the one-loop calculation of $M_\pi$ and $F_\pi$ gives
\cite{GLFV}:
\bea
M_\pi(L)&=& M_\pi \left[1+\frac{1}{2 N_f}\xi\, g_1(\lambda)+O(\xi^2)\right]
\nonumber
\\
F_\pi(L)&=&\;F_\pi\left[1-\;\frac{N_f}{2}\,\xi\, g_1(\lambda)+O(\xi^2)\right] 
\label{eq:FM1L}
\eea
with $\lambda=M_\pi L$, $\xi=(M_\pi/4 \pi F_\pi)^2$ and
\bea
g_1(\lambda) &=& \sum_{\vec{n}} \nolimits' \int_0^\infty dz \;
e^{-{ 1 \over z}-\frac{z}{4} \vec{n}^2 \lambda^2} \nonumber \\
&=& \sum_{k=1}^\infty \frac{4
  m(k)}{\sqrt{k} \lambda} K_1(\sqrt{k} \lambda) 
\co
\label{eq:g1}
\eea
is the tadpole correction in finite volume ($m(k)$ is the multiplicity of a
vector $\vec{n}$ with $\vec{n}^2=k$). The result shows 
that finite volume corrections for masses and decay constants are
suppressed exponentially, by increasing powers of $\exp(-M_\pi L)$. The
infinite sum appearing in Eq.~(\ref{eq:g1}) reflects the infinite sum in
the definition of a fully periodic propagator, Eq.~(\ref{eq:GL}), but since
$M_\pi L \gg 1$ only the first few will be numerically relevant.

A different approach which relies on an expansion in exponentials and not
on CHPT is due to L\"uscher \cite{luscher} -- this expansion, however, has
only been devised for masses and, more recently, for decay constants
\cite{CH}, and is therefore less general than the chiral expansion. A
detailed comparison of the two approaches will be given later.

\subsection{Overview of recent results}
The examples just discussed show that, in the $p$--regime finite-volume
effects are typically small and relevant for very precise lattice
calculations. Indeed only recently there has been an intense activity in
analytical calculations of these effects despite the fact that the
theoretical framework had been established around twenty years ago
\cite{GLFV,luscher}, and extended to the quenched approximation not much
later \cite{BG,sharpe}.

During last year there have been analyses of finite volume effects for:
two-pion state energies \cite{Lin:2003tn}, the pion mass \cite{CD}, $F_K$
and $B_K$ \cite{BV}, the nucleon mass \cite{QCDSF,koma}, the nucleon
magnetic moments and axial coupling constant as well as the mass
\cite{beane}, $f_B$ and $B_B$ \cite{arndt} and $\fpi$ \cite{CH}. Many of
these results have been also presented at this conference
\cite{lin-lat04,goeckeler,koma-lat04} together with new ones concerning
$M_\pi, \; F_\pi$ and the charge radius \cite{lewis}. Most of these
calculations have been made in the framework of CHPT and concerned one-loop
finite-volume effects and not using L\"uscher formula. A comparison to the
L\"uscher formula has been made in the case of the proton mass
\cite{QCDSF}, and a discrepancy has been found with the explicit formula
given by L\"uscher in his Carg\`ese lecture notes \cite{cluscher}. It turns
out that the formula in these lecture notes was not correctly written and
that the correction has numerically important effects
(cf.~\cite{QCDSF,koma}).

In what follows I will discuss in some more detail why it is important to
go beyond the leading-order calculation of these effects if one wants to
have a good control over them, and how this can be achieved by a combined
use of CHPT and the asymptotic formulae {\em \`a la} L\"uscher. This is the
approach advocated in \cite{CD,CH}. 

\subsection{Hadron masses in finite volume: CHPT vs. L\"uscher Formula}
Above we have briefly discussed what rules one has to follow if one wants
to calculate finite volume corrections to a given quantity in CHPT, and
shown the result for the pion mass. The analogous result for the kaon, 
the proton or any other hadron mass will of course be different. L\"uscher,
however, has shown that the leading exponential correction to the 
mass of a given hadron can be expressed in a completely universal way through
an integral over the scattering amplitude of the hadron in question and the
pion. The latter enters here because it is the lightest hadron: the leading
exponential correction to any hadron mass is of the form $\exp(-M_\pi L)$,
and how large the correction is depends on how strong is the interaction of
the hadron in question to the pion. For example the L\"uscher formula
for the pion mass reads
\bea
\label{eq:mpiL}
M_\pi(L)-M_\pi =-\frac{3}{16\pi^2 M_\pi L}\times \hskip 2 cm && \\ 
\int_{-\infty}^\infty\!dy\; 
F({\rm i}y)\,e^{-\sqrt{M_\pi^2+y^2}\,L}+O(e^{-\overline{M}L}) &&
\nonumber
\eea
where $F(\nu)$ is the physical (for $\nu$ real -- what enters the integral
is the analytic continuation) forward $\pi \pi$ scattering amplitude and
$\overline{M} > \sqrt{2} M_\pi$. Note that the formula does not rely in any
way on the chiral expansion, and indeed is valid also in cases where there
is no spontaneous symmetry breaking and the corresponding Goldstone bosons
-- in such cases the role of the pion has to be overtaken by the particle
which happens to be the lightest one. 

The formula can be used for a numerical analysis only if one has a
representation for the scattering amplitude in the integrand which lends
itself to a numerical evaluation. For hadrons one could in principle take
directly the measured scattering amplitude (and do the necessary analytic
continuation) to evaluate the integral, but then one would be restricted to
physical values of the quark masses: in order to use the formula for
arbitrary values of the quark masses one must therefore rely on the chiral
representation for the scattering amplitude. Inserting the leading order
chiral representation for the $\pi \pi$ scattering amplitude:
\be
F(\nu) = -\frac{M_\pi^2}{F_\pi^2}+O(M_\pi^4)
\label{eq:F2}
\ee
in Eq.~(\ref{eq:mpiL}) one should obtain the same expression for the
leading exponential term in Eq.~(\ref{eq:FM1L}). As was verified long ago
\cite{GLFV}, this is indeed the case.

\begin{figure}[t]
\includegraphics[width=75mm]{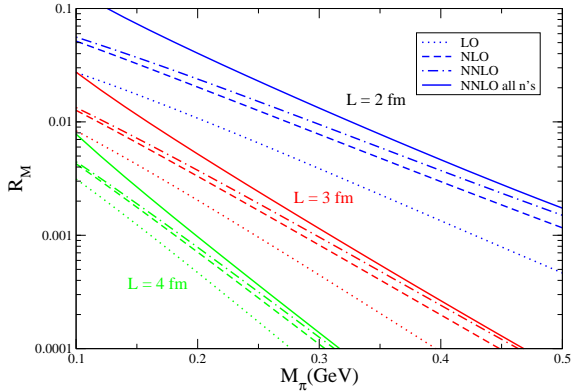}
\caption{Relative finite volume corrections ($R_M=(M_{\pi,L}-M_\pi)/M_\pi$)
  to the pion mass using the L\"uscher formula (\ref{eq:mpiL}) or the
  modified version which includes subleading exponentials (\ref{eq:MLF}) and
  the chiral expansion of the $\pi \pi$ amplitude up to NNLO. 
\label{fig:mpi}}
\end{figure}
The chiral expansion for the scattering amplitude is now known up to and
including order $p^6$ \cite{BCEGS}, and it is easy to insert the
corresponding analytic expressions in Eq.~(\ref{eq:mpiL}) and evaluate the
integral. This has been done in \cite{CD}, and the corresponding numerical
results are shown in Fig.~\ref{fig:mpi}. A somewhat surprising feature of
the numerical results is that the NLO corrections are quite substantial and
of the order of 50\% of the LO effect. The comparison of NLO and NNLO
however shows that chiral expansion does have a good converging behaviour
for all values of $M_\pi$ and $L$ in the region of applicability of the
formula. 

A comparison to the numerical values obtained with the full one-loop CHPT
formula Eq.~(\ref{eq:FM1L}) on the other hand shows that for not so large
values of $M_\pi L$ exponentially subleading terms may be important. As was
suggested in \cite{CD} the most reliable estimate of the finite volume
effects for the pion mass is obtained by adding to the full one-loop CHPT
result the NLO and NNLO corrections obtained through the L\"uscher formula.
There is actually a simple extension of the L\"uscher formula that provides
a convenient combination of both approaches. The modified formula is the
following: 
\bea
\label{eq:MLF}
M_{\pi,L}-M_\pi = - \frac{1}{32 \pi^2 \lambda} \sum_{k=1}^\infty
{m(k) \over \sqrt{k}} \times && \hskip 0.5 cm \\
\int_{- \infty}^\infty dy F(i y)
e^{-\sqrt{k (M_\pi^2 + y^2)}L}  +O(e^{-\overline{M}L}) && \nonumber
\eea
and can be derived with the following reasoning. In L\"uscher's paper
\cite{luscher} a substantial part of the proof of the formula is devoted to
showing that the leading exponential corrections are obtained from loop
graphs where only one pion propagator is taken in finite volume (i.e. is
made periodic, as in Eq.~(\ref{eq:GL})) and all other propagators are
the standard, infinite-volume ones. Since he concentrates only on the
leading exponential correction, he then immediately drops all terms with
$|\vec{n}|>1$ in the infinite sum which makes the propagator periodic. But
nothing forbids to keep all terms in the infinite sum Eq.~(\ref{eq:GL}) and
to follow the rest of the derivation which leads to Eq.~(\ref{eq:MLF}).

A few comments are in order:
\begin{enumerate}
\item 
although Eq.~(\ref{eq:MLF}) contains exponentially subleading terms, it is
not algebraically better than Eq.~(\ref{eq:mpiL}) because both give only
the leading term {\em exactly} -- numerically, however, the effect of
subleading terms is important for moderate values of $M_\pi L$ as shown in
Fig.~\ref{fig:mpi};
\item
an inclusion of exponentially subleading terms which is similar in spirit
to the one proposed here has been already introduced by L\"uscher by
extending the integration boundaries to infinity in Eq.~(\ref{eq:mpiL});
\item
at the two-loop level there will be two types of contributions which are
not included in Eq.~(\ref{eq:MLF}): contributions where two pion
propagators are taken in finite volume, and contributions to the loop
integration from other singularities other than the pion pole in the
propagator (note that both contributions are absent at the one-loop level:
indeed if one inserts the leading order amplitude (\ref{eq:F2}) in
Eq.~(\ref{eq:MLF}) one recovers exactly the full one-loop result
(\ref{eq:FM1L})). Only a complete two-loop calculation will show how big
these contributions are numerically \cite{CH2}.
\end{enumerate}

\subsection{Decay constants in finite volume}
The discussion in the previous subsection has shown the usefulness of the
asymptotic formula of L\"uscher in evaluating finite-volume corrections for
masses, and calls for extensions to other quantities. Recently, an
extension to decay constants has been provided in \cite{CH} and reads
\bea
F_{\pi,L}-F_\pi = \frac{3}{8 \pi^2 M_\pi L} \times && \nonumber \\ 
\int_{-\infty}^{\infty}d y \; e^{-\sqrt{\mpi^2+y^2} L} N_F(iy) +O(e^{-\bar
  ML}) \co &&
\label{eq:Fpi}
\eea
where $N_F(\nu)$ is (as one intuitively expects) related to the 
$\langle 0|A_\mu| 3 \pi \rangle$ amplitude. In this case there is a
subtlety related to the fact that the latter amplitude has a pole due to
the direct coupling of the axial current to a pion, which then rescatters
into three pions. This pole appears exactly in the kinematical region where
it is needed in Eq.~(\ref{eq:Fpi}) and must be subtracted. The prescription
for the subtraction of this pole has been discussed in detail in \cite{CH},
together with the physical interpretation of the subtraction.

\begin{figure}[t]
\includegraphics[width=75mm]{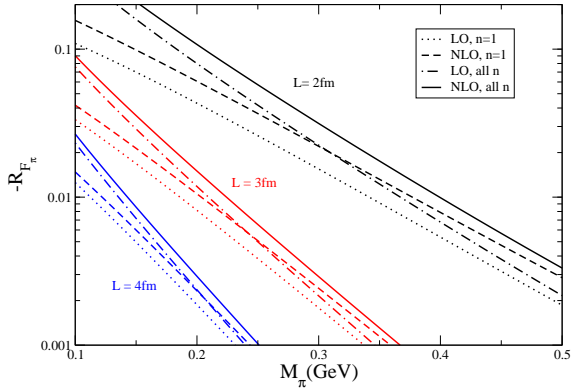}
\caption{Relative finite volume corrections to the pion decay constant
  ($R_{F_\pi}=(F_{\pi,L}-F_\pi)/F_\pi$). 
\label{fig:fpi}}
\end{figure}
This asymptotic formula can now be used for a numerical analysis in the
same way as the L\"uscher formula for the pion mass: in the present case
the needed infinite-volume amplitude is the one relevant for the decay of
the $\tau$ into a neutrino and three pions and has been calculated to one
loop in CHPT \cite{tau3pi}. The numerical results, including the effects of
subleading exponential terms, included according to the analogous of
Eq.~(\ref{eq:MLF}) are shown in Fig.~\ref{fig:fpi}. 
\begin{figure}[t]
\includegraphics[width=75mm]{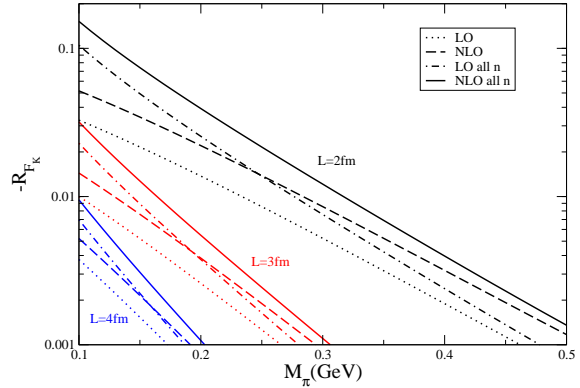}
\caption{Relative finite volume corrections to the kaon decay constant
($R_{F_K}=(F_{K,L}-F_K)/F_K$). \label{fig:fk}}
\end{figure}
Like L\"uscher's formula, this extension to decay constants can also
be applied to heavier hadrons: what determines the finite-volume correction
to the decay constant of the hadron H is the amplitude $\langle 0|A_\mu| 2
\pi H \rangle$. For the case of the kaon, e.g., this is the amplitude
relevant for $K_{l4}$ decays, which is nowadays known to two loops
\cite{ABT}. A numerical analysis based on the one-loop evaluation of this
amplitude \cite{KL41} gives the results shown in Fig.~\ref{fig:fk}. 
Finite volume corrections for $F_\pi$ and $F_K$ are also of
phenomenological interest, because the lattice calculation of these two
quantities can be used to determine $V_{us}$, as suggested by Marciano
\cite{marciano}, and performed by the MILC collaboration \cite{MILC}. 
A complete numerical analysis of the masses and decay
constants of all the octet of pseudoscalars is in progress \cite{CH2}.

\section{Conclusions}
There has been a lot of activity recently in calculating finite volume
effects analytically. I have briefly reviewed the theoretical tools needed
for performing these calculations as well as some of this recent activity.
The cost of a lattice calculation grows very rapidly with the box size --
the calculations discussed here show that in many cases it is unnecessary
to make the infinite-volume extrapolation numerically, and one can correct
for finite-volume effects analytically.

\section*{Acknowledgments}
It is a pleasure to thank the organizers for the invitation and the perfect
organization of the conference, Leonardo Giusti, Karl Jansen and Hartmut
Wittig for informative discussions, and Stephan D\"urr, Andreas Fuhrer and
Christoph Haefeli for a pleasant and frutiful collaboration. Christoph
also for a careful reading of the manuscript. Work supported by
Schweizerische Nationalfonds and in part by RTN, BBW-Contract No. 01.0357
and EC-Contract HPRN--CT2002--00311 (EURIDICE).

\end{document}